\documentclass[aps,amsmath,reprint,prx]{revtex4-2}
\RequirePackage{url}
\RequirePackage{natbib}
\RequirePackage{ifthen}
\RequirePackage[utf8]{inputenc}
\RequirePackage{newunicodechar} 

\newunicodechar{ }{ }  % This particularly mysterious command tells LaTeX that unicode character U+2009 is, in fact, a space
\newunicodechar{−}{\ensuremath{-}}        % unicode minus sign.  Not the same as a hyphen, please.
\newunicodechar{∕}{\ensuremath{/}}        % unicode divide: note Math division as opposed to text solidus
\newunicodechar{×}{\ensuremath{\times}}   % unicode times
\newunicodechar{°}{\degree}               % unicode degree sign, \degree is from gensymb

\newboolean{ShowComments}
\setboolean{ShowComments}{false}  % change to true to show remaining colorful author comments. 
\input{frontmatter}

\begin{document}
\renewcommand\graphicspath[1]{./figures/#1}
\title{Full-permutation dynamical decoupling in triple-quantum-dot spin qubits}

\author{Bo~Sun} \email[]{bsun@hrl.com}
\author{Teresa~Brecht}
\author{Bryan~Fong}
\author{Moonmoon~Akmal}
\author{Jacob~Z.~Blumoff}
\author{Tyler~A.~Cain}
\author{Faustin~W.~Carter}
\author{Dylan~H.~Finestone}
\author{Micha~N.~Fireman}
\author{Wonill~Ha}
\author{Anthony~T.~Hatke}
\author{Ryan~M.~Hickey}
\author{Clayton~A.~C.~Jackson}
\author{Ian~Jenkins}
\author{Aaron~M.~Jones}
\author{Andrew Pan}
\author{Daniel~R.~Ward}
\author{Aaron~J.~Weinstein}
\author{Samuel~J.~Whiteley}
\author{Parker~Williams}
\author{Matthew~G.~Borselli}
\author{Matthew~T.~Rakher}
\author{Thaddeus~D.~Ladd}

\affiliation{HRL Laboratories, LLC,
3011 Malibu Canyon Rd., Malibu, CA 90265}
\date{\today}

\begin{abstract}
Dynamical decoupling of spin qubits in silicon can enhance fidelity and be used to extract the frequency spectra of noise processes. 
We demonstrate a full-permutation dynamical decoupling technique that cyclically exchanges the spins in a triple-dot qubit. 
This sequence not only suppresses both low frequency charge-noise- and magnetic-noise-induced errors; it also refocuses leakage errors to first order, which is particularly interesting for encoded exchange-only qubits. 
For a specific construction, which we call NZ1y, the qubit is isolated from error sources to such a degree that we measure a remarkable exchange pulse error of $5\times10^{-5}$.
This sequence maintains a quantum state for roughly 18,000 exchange pulses, extending the qubit coherence from $T_2^*=2~\mu$s to $T_2 = 720~\mu$s.
We experimentally validate an error model that includes $1/f$ charge noise and $1/f$ magnetic noise in two ways: by direct exchange-qubit simulation, and by integration of the assumed noise spectra with derived filter functions, both of which reproduce the measured error and leakage with respect to changing the repetition rate. 

\end{abstract}
\maketitle

\newcommand{\tp}{\ts{t}{pulse}}
\newcommand{\ti}{\ts{t}{idle}}

\section{Introduction}
\storyboard{(Introduction and references other DD papers..)}
Dynamical decoupling (DD) sequences suppress dephasing in quantum systems by periodically inverting interactions between the qubit and its environment~\cite{Viola98,Ezzel2022}.  
Applied to qubits based on electron spins in silicon, DD can extend qubit coherence times to more than a second in donor-bound spins \cite{Tyryshkin2012} and more than 20~ms in quantum dots \cite{Veldhorst2014}. 
For nuclei in silicon, dynamically decoupled coherence times have been shown to be at least hours long for ensembles~\cite{Saeedi2013} and more than 30~s for a single nucleus~\cite{Muhonen2014}. 
While extended qubit memory is one motivation of DD experiments, the present work focuses on another key function: DD can expose features of noise processes relevant to qubit performance in quantum information processing systems.
Specifically, periodic DD sequences act as frequency domain filters applied to the noise spectrum witnessed by spins. 
By using this filter-function formalism, time domain DD data can be inverted to extract frequency domain noise spectra 
\cite{Bylander2011,Alvarez2011,Kerckhoff2021,Kawakami2016,Yoneda2018,Struck2020,Nakajima2020,Connors2022}.

\storyboard{(Define full-permutation dynamical decoupling.)}
A standard DD process is the famous Hahn spin-echo~\cite{Hahn1950}, in which the effective interaction between a spin and its local magnetic field is periodically reversed by application of spin-flipping $\pi$ pulses. 
In this work, we demonstrate a somewhat different application in our system, beginning by preparing three spins into a Decoherence Free Subsystem (DFS), for which the total spin projection is a degree of freedom that is removed from initialization, control, and read-out~\cite{Knill2000,DiVincenzo2000,Andrews2019}. 
The DFS is insensitive to global magnetic fields that couple to the total spin projection, but sensitive to local sources of noise: magnetic field gradients and charge noise.
Our decoupling sequence then relies on the principle of periodically and symmetrically permuting the three spins, which causes local noise to average into a global term to which the encoded DFS is impervious~\cite{Wu2002,West2012}.
We refer to this process as full-permutation dynamical decoupling.
Unlike electron shuttling techniques \cite{Mortemousque2021}, this form of DD exchanges only the electron spin state, and is particularly well suited to exchange-only qubits as the DFS subspace also serves as a basis for encoded universal control using the voltage-controlled exchange interaction~\cite{DiVincenzo2000,Fong2011,Maune2012,Eng2015,Reed2016,Jones2018,Andrews2019}.
In this paper, we demonstrate that full-permutation DD can supress error rates to $5\times10^{-5}$ per control pulse, which is an improvement of two orders of magnitude compared to the error rate of the same qubit obtained from randomized benchmarking ($1.3\times10^{-3}$ per exchange pulse).
Furthermore, we show that this decoupling sequence can be used to verify our understanding of qubit noise sources by comparing experimental results with simulation across regimes of magnetic noise and charge noise dominance accessed by varying the repetition rate.
We show that the sequence we employ is robust to low frequency noise as well as miscalibration, and it therefore allows us to validate a $1/f$ spectral character of each of these noise sources to greater precision than is possible using methods like randomized benchmarking, which elucidates design trade-offs for future device iterations.

\storyboard{(Structure of paper)}
We begin by describing our triple-dot qubit in \refsec{sec:qubit}.
We will then detail the structure and noise filtering capability of full-permutation DD in \refsec{sec:DD}, followed by the experimental results in \refsec{sec:results}.
As we will discuss, this decoupling strategy is highly effective at eliminating dephasing due to local magnetic fields that vary slowly compared to the time scale of the qubit control pulses.
This provides a key gauge of operational noise for the DFS system.

\section{The $\text{Si/SiGe}$ Exchange-Only Triple-Dot Qubit}\label{sec:qubit}
\storyboard{(We will focus on SiGe.)}
In this paper, we explore the performance of a full-permutation DD sequence applied to an exchange-only triple-dot qubit within an isotopically enhanced silicon quantum well~\cite{Borselli2015,Eng2015,Reed2016,Jones2019}.
The quantum dots are formed by the electrostatic potential created by patterned metal gates on a SiGe/$^{28}$Si/SiGe heterostructure, in which the quantum well is 3~nm thick and the $^{29}$Si content is reduced to 800~ppm.
In contrast to recent devices fabricated using the single-layer etch-defined gate electrode (SLEDGE) technique~\cite{Ha2022}, this device uses an Al overlapping gate design ~\cite{Zajac2015,Borselli2015,Reed2016}.
A false-color scanning electron micrograph (SEM) of a representative device is shown in Fig.~\ref{FIG:dev_seq}b, where the plunger gate (P) voltages are adjusted to trap a single electron, and the exchange interaction between neighboring electrons is controlled by voltages applied to the exchange gates (X).
Readout is achieved using charge sensors (M) and a signal chain described in Ref.~\onlinecite{Blumoff2021}.
The device geometry and methods of calibration and control are the same as in Ref.~\onlinecite{Andrews2019}.

\storyboard{(specifically, triple dot qubits. It's encoded and there exists a gauge freedom.)}
In this device, we focus on gates P1, P2, and P3, each of which trap a single spin.
The 8 basis states for these three spins may be written as $\ket{S_{12},S;m}$, where $S=1/2,3/2$ is the total spin quantum number across all three dots corresponding to $\vec{S}=\vec{S}_1+\vec{S}_2+\vec{S}_3$, $m=-S,1-S\ldots,S$ is the total spin projection, and $S_{12}$ is the combined spin of the electrons in dots 1 and 2, corresponding to $\vec{S}_{12}=\vec{S}_1+\vec{S}_2$.
This assignment of spin quantum numbers is shown in cartoon form in Fig.~\ref{FIG:dev_seq}f and discussed in Ref.~\cite{Burkard2021}.
The encoded DFS qubit lives in the subsystem where $S=1/2$.
Encoded $\ket{0}=\ket{0,1/2;m}$ is the singlet state of the first two spins, and encoded $\ket{1}=\ket{1,1/2;m}$ is a superposition of the triplet states of the first two spins; both of these ``states" are in fact doublets for $m=\pm 1/2$.
A random selection of the two $\ket{0}$ states is initialized by preparing the first two spins in a singlet ground state, leaving the third spin unpolarized~\cite{Eng2015}, i.e. $\rho_0 = \sum_{m=\pm 1/2} \ketbra{0,1/2;m}{0,1/2;m}/2$.
The $S_{12}$ quantum number is directly read out via singlet-triplet readout employing Pauli spin blockade, which provides no information about $S$ or $m$.
The four states $\ket{1,3/2;m}$ (which we refer to collectively as $\ket{Q}$, for quadruplet), are outside of the DFS and referred to as leaked states.
Since all $\ket{Q}$ states have $S_{12}=1$, they appear as encoded $\ket{1}$ during measurement.  

\storyboard{(Explain the qubit and it's control axes. Encoded qubit is insensitive to global fields, but not gradients.)}
This qubit is universally controllable via exchange interactions alone~\cite{DiVincenzo2000,Fong2011,Andrews2019,Burkard2021}, i.e. with the control Hamiltonian
\begin{equation}
	\label{eq:Hcontrol}
	\ts{H}{control}(t)=J_{12}(t)\vec{S}_1\cdot\vec{S}_2 + J_{23}(t)\vec{S}_2\cdot\vec{S}_3,
\end{equation}
where $\vec{S}_j$ is the single-spin operator for spin $j$ and $J_{jk}(t)$ is the time-varying exchange energy between spins $j$ and $k$.
Each exchange interaction is activated by pulsing the voltage of an X gate above a tunnel barrier between dots; enhanced tunneling reduces the energy of the singlet-state between those two spins for the duration of the pulse.
The symmetry of this exchange-only Hamiltonian means it has no impact on quantum numbers $S$ or $m$; it only impacts the $S_{12}$ quantum number.
For the single qubit we study here, we may therefore provide a geometric analogy for exchange gates.
In the Bloch sphere representation of the DFS, Fig.~\ref{FIG:dev_seq}d and e, exchange between dots 1 and 2 drives rotations about the $\hat{z}$-axis, and exchange between dots 2 and 3 drives rotations about the $\hat{n}$-axis, which is $120^\circ$ separated from $\hat{z}$ in the $x$-$z$ plane (according to Clebsch-Gordan coefficients.)
Full control is therefore possible using calibrated exchange pulses on these axes.
For full-permutational DD, the only calibrated exchange pulse needed is the $\pi$ pulse, which enacts a full spin swap.
A $\pi$-pulse about the $\hat{z}$-axis (whose unitary, a Pauli operator, is abbreviated $Z$) is in fact a swap of spins 1 and 2.
A $\pi$-pulse about the $\hat{n}$-axis (whose unitary is abbreviated $N$) is a swap of spins 2 and 3.

\storyboard{(Magnetic noise; gradients exist and hurt.)}
Here, we classify noise in our system as coming from one of two forms.
First is magnetic field noise, described by the Hamiltonian
\begin{equation}
	{H}_{B}=-g\mu_B\sum_{j} \vec{B}_j(t) \cdot \vec{S}_j + \sum_{jk} A_{jk} \vec{I}_k\cdot\vec{S}_j.
	\label{eq:HB}
\end{equation}
where $g$ is the gyromagnetic ratio ($\approx 2$ in Si), $\mu_B$ the Bohr magneton, $\vec{B}_{j}$ the magnetic field at dot $j$, $\vec{I}_k$ the spin of the $k$th nuclear spin ($^{29}$Si or $^{73}$Ge) in the Si/SiGe heterostructure, and $A_{jk}$ is the contact hyperfine coupling energy, proportional to the probability of finding spin $j$ at the location of nucleus $k$~\cite{Kerckhoff2021}.
For simplicity, define an effective-field angular frequency $\vec{b}_j=(-g\mu_B \vec{B}_j + \sum_k A_{jk}\vec{I}_k)/\hbar$.
Our DFS encoding assures that neither a static value nor random fluctuations in $\sum_j \vec{b}_j(t)$ have any measurable effect on our qubit, since global fluctuating magnetic fields couple only to the $\vec{S}$ degree of freedom, which impacts only $m$, but not our qubit degree of freedom $S_{12}$.
However, gradients such as $\vec{b}_1-\vec{b}_2$ do not conserve the $S_{12},S,$ and $m$ quantum numbers, and thus cause decoherence by both impacting subsystem qubit states and causing leakage out of the DFS.
Such gradients are particularly important due to the spatially-varying \Si\ and \Ge\ nuclear spin baths present in our devices.

\storyboard{(Pulse errors exist but don't leak.)}
A second form of relevant noise comes from imperfect exchange operations.
When we attempt a $\pi$ pulse of duration $\ts{t}{pulse}$ on spins 2 and 3, the integrated angle is
\begin{equation}
	\frac{1}{\hbar}\int_t^{t+\ts{t}{pulse}} (J_{23}(\tau) +\delta J_{23}(\tau)) d\tau = \pi + \delta\theta_n(t),
\end{equation}
where the noise $\delta J_{23}(t)$ may result from various sources including miscalibration, local charge noise,  and noise from control instruments.
As a result, when we attempt to perfectly swap spins $2$ and $3$, we deviate from this ideal by $\exp(-i\delta\theta_n \vec{S}_2\cdot\vec{S}_3)$, interpretable as over- or under-rotation about the $\hat{n}$-axis.
Likewise, noise on $J_{12}(t)$ causes an integrated angle deviation of $\delta\theta_z(t)$.
Such deviations cause qubit error, but they conserve $S_{12}$ and $m$ and consequently do not cause leakage.

\section{Full-Permutation Dynamical Decoupling}\label{sec:DD}

\begin{figure}[th!]
	\includegraphics[width=\columnwidth]{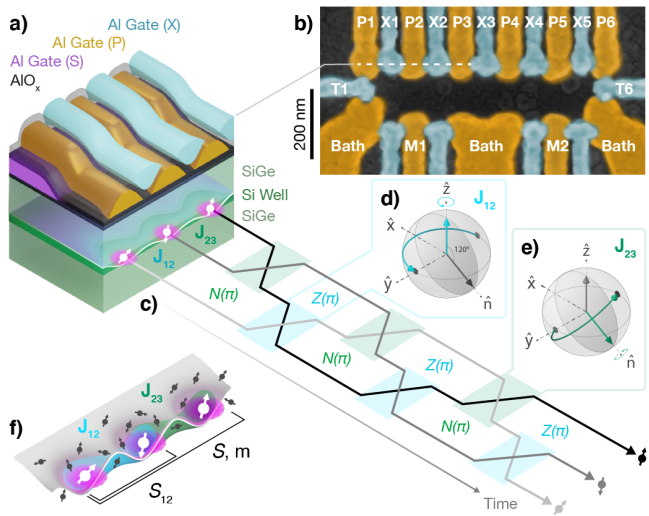}
	\caption{Full-permutation DD (NZ sequence) in a triple-dot exchange-only qubit. 
		\textbf{(a)} 3D sketch of the Si/SiGe heterostructure and gate stack, including three electron spins occupying dots in the quantum well confined laterally by the gate stack above.  
		\textbf{(b)} A false-color scanning electron micrograph with plunger gates (P), exchange gates (X), bath tunnel barrier gates (T) and electrometer gates (M) labeled. 
		\textbf{(c)} Illustration of the decoupling sequence.
		\textbf{(d)} Bloch sphere showing  $\pi$ rotation on $J_{23}$ (or $N$ operation) resulting from a pulse on the X2 gate.
		\textbf{(e)} Bloch sphere showing  $\pi$ rotation on $J_{12}$ (or $Z$ operation) resulting from a pulse on the X1 gate.
		\textbf{(f)} Illustration showing randomly-oriented spinful nucei of the lattice (grey arrows), the slow fluctuations of which may be dynamically decoupled.
	} 
	\label{FIG:dev_seq}
\end{figure}

The theory of full-permutation DD was established in Ref.~\onlinecite{Wu2002}, elaborated by Ref.~\onlinecite{West2012} to include the development of higher-order sequences than we apply here.
Full-permutation DD homogenizes the gradient magnetic field across the three spins by successively swapping pairs of spins in the triple-dot qubit.
In the Bloch-sphere picture for the exchange-only qubit, the sequence applies alternating rotations around the $\hat{n}$-axis and $\hat{z}$-axis, and hence we use a shorthand to refer to full-permutation DD as an ``NZ" sequence.
The NZ sequence is illustrated by the braid in Fig.~\ref{FIG:dev_seq}c, which shows that over the duration of 3 repetitions of $NZ$, each spin spends an equal amount of time localized within each quantum dot before returning to its initial position.
Thus, the six pulse sequence $NZNZNZ$ forms the base decoupling block.  
As this is a first-order sequence relative to the constructions in Ref.~\onlinecite{West2012}, we refer to it as NZ1.

Analogously to the analysis of the famous Carr-Purcell-Meiboom-Gill (CPMG) sequence developed half a century ago in the context of nuclear magnetic resonance~\cite{Carr1954,Meiboom1958}, the operation of the NZ1 sequence on qubit states in the presence of low-frequency noise and static inhomogeneities may be understood via average Hamiltonian theory.
For this, we consider our total noise Hamiltonian discussed in \refsec{sec:qubit},
\begin{multline}
	\label{eq:Hnoise}
	\ts{H}{noise}(t)=\\
	\sum_j \vec{b}_j(t)\cdot\vec{S}_j 
				+ \delta J_{12}(t) \vec{S}_1\cdot\vec{S}_2
				+ \delta J_{23}(t) \vec{S}_2\cdot\vec{S}_3,
\end{multline}
modulated by the ``toggling frame" or interaction picture: 
\begin{equation}
\ts{\tilde{H}}{noise}(t)=U_{\text{NZ}1}^\dag(t) \ts{H}{noise}(t) U_{\text{NZ1}}^{\phantom\dag}(t).
\end{equation}
Here $U_{\text{NZ1}}(t)$ is the periodic unitary corresponding to applying exchange swaps, alternating between $N$ and $Z$ square pulses of duration $\tp$ interspersed by periods of duration $\ti$, where exchange is negligible.  This unitary is straightforward to calculate for any exchange pulse shape $J_{jk}(t)$, as long as the pulses for $J_{12}(t)$ and $J_{23}(t)$ never overlap in time.
If the time-dependence of \refeq{eq:Hnoise} is either very slow relative to the period of $U_{\text{NZ1}}(t)$ or has the same periodicity (as in the $\delta J_{jk}(t)$ terms), then Floquet's theorem assures that $\ts{\tilde{H}}{noise}(t)$ generates a unitary of the form $U_p(t)\exp(-iFt)$, where $U_p(t)$ has the same periodicity as $U_{\text{NZ1}}(t)$ (so $U_p(nT)=1$ for integer $n$ and period $T$), and $F$ is the Floquet Hamiltonian.
If measured at each full period of six pulses, this Floquet Hamiltonian captures the slow dynamics of the spins due to $\ts{H}{noise}$.
It may be calculated perturbatively using the Magnus expansion over a full period of $U_{\text{NZ1}}(t)$ as $F=\sum_n\bar{H}^{(n)}$, with each term $n$th order in $\ts{H}{noise}$.
For the present analysis of NZ1, we will consider only the lowest order term:
\begin{equation}	
	\bar{H}^{(0)} = \frac{1}{T}\int_{0}^{T} d\tau  \ts{\tilde{H}}{noise}(\tau).
\end{equation}

\com{
To unpack this notation, we provide the example that the component of this averaging integral between times $\tp+\ti$ and $2\tp+\ti$, corresponding to the finite-width application of the second pulse, a $Z$ pulse, of each $NZNZNZ$ period, is
\begin{multline}
\int_{nT+\tp+\ti}^{nT+2\tp+\ti} d\tau \ts{\tilde{H}}{noise}(\tau) = 
\\
\int_{0}^{\tp} N e^{-i\vec{S}_1\cdot\vec{S}_2\pi\tau/\tp}  
	\ts{H}{noise} e^{i\vec{S}_1\cdot\vec{S}_2\pi\tau/\tp}N  d\tau
\\
=
\int_{0}^{\tp} N \biggl[
\cos^2\biggl(\frac{\pi\tau}{\tp}\biggr) 
	(\vec{b}_1\cdot\vec{S}_1+\vec{b}_2\cdot\vec{S}_2)
\\
\phantom{+\int_{0}^{\tp} N}+
\sin^2\biggl(\frac{\pi\tau}{\tp}\biggr) 
	(\vec{b}_1\cdot\vec{S}_2+\vec{b}_2\cdot\vec{S}_1)
\\
\phantom{+\int_{0}^{\tp} N}-
\sin\biggl(\frac{\pi\tau}{\tp}\biggr) (\vec{b}_2-\vec{b}_1)\cdot\vec{S}_1\times\vec{S}_2
\\
+\vec{b}_3\cdot\vec{S}_3
+\delta J_{12}\vec{S}_1\cdot \vec{S}_2
\biggr]N  
\\
% =
% N \biggl[
% \frac{\tp}{2}(\vec{b}_1+\vec{b}_2)\cdot(\vec{S}_1+\vec{S}_2)
% \\
% -\frac{2\tp}{\pi} (\vec{b}_1-\vec{b}_1)\cdot\vec{S}_1\times\vec{S}_2
% \\
% +\tp \vec{b}_3\cdot\vec{S}_3
% +\delta\theta_z \vec{S}_1\cdot \vec{S}_2
% \biggr]N  d\tau
% \\
=
\frac{\tp}{2}(\vec{b}_1+\vec{b}_2)\cdot(\vec{S}_1+\vec{S}_3)
-\frac{2\tp}{\pi} (\vec{b}_1-\vec{b}_1)\cdot\vec{S}_1\times\vec{S}_3
\\
+\tp \vec{b}_3\cdot\vec{S}_2
+\delta\theta_z \vec{S}_1\cdot \vec{S}_3.
\end{multline}
By summing such terms for each such pulse and idle, we arrive at the average Hamiltonian
\begin{multline}
	\bar{H}^{(0)} = \biggl[\frac{\tp}{T}\vec{G}+2\biggl(\tp +\ti\biggr)\sum_j \vec{b}_j\biggr]\cdot\vec{S}
	\\
	+\frac{1}{T}(\delta\theta_z+\delta\theta_n)\biggl(\sum_{j\ne k}\vec{S}_j\cdot\vec{S}_k\biggr)
	\\
	+\frac{2\tp}{\pi T}\vec{G}\cdot[(\vec{S}_1+\vec{S}_3)\times\vec{S}_2-\vec{S}_1\times\vec{S}_3],
\label{aveH}
\end{multline}
where 
\begin{equation}
	\vec{G} = \frac{\vec{b}_1+\vec{b}_3}{2}-\vec{b}_2.
\end{equation}
\purple{(Discuss each term.)}

The first line of Eq.~\eqref{aveH} is the one intuitively illustrated in Fig.~\ref{FIG:dev_seq}:  Local fields average to a global field which couples to the operator $\vec{S}$, impacting only the ignored $m$ quantum number.
The second line is also notable: It says that a random but static miscalibration of $\pi$ pulses couples at lowest order to all three exchange-pairings in equal magnitude, but $\sum_{j\ne k}\vec{S}_j\cdot\vec{S}_k = S(S+1)-9/4$, which results only in an overall phase between the encoded and leakage subsystems; hence NZ1 is tolerant to static miscalibration to first order.
}

The result of this calculation (using period $T=6\ts{t}{pulse}+6\ts{t}{idle}$) is
\begin{multline}
	\bar{H}^{(0)} = \frac{\delta\theta_n+\delta\theta_z}{16\tp+16\ti}(-1)^{S-1/2}
	+\frac{\vec{b}_1+\vec{b}_2+\vec{b}_3}{3}\cdot\vec{S}\\
	+\frac{\tp}{12\tp+12\ti}\biggl\{
	[(\vec{b}_2-\vec{b}_3)C_{23}+(\vec{b}_2-\vec{b}_1)C_{12}]\cdot\\
	[\vec{S}_1\times\vec{S}_2+\vec{S}_2\times\vec{S}_3
	+\vec{S}_3\times\vec{S}_1]\biggr\},
\end{multline}
where 
\be
C_{jk}   = \frac{1}{2\tp}\int_0^{\tp} \sin\biggl(\frac{1}{\hbar}\int_0^t J_{jk}(t')dt'\biggr)dt.
\ee
From this expression, the exchange noise terms (i.e. containing $\delta\theta_\alpha$) provide only a phase shift to leakage spaces, and the average magnetic field $\sum_k \vec{b}_k/3$ couples only to $\vec{S}$, which impacts only $m$. 
Notably, neither of these effects are detectable or impactful on the encoded qubit.
The remaining terms depend on magnetic gradients $\vec{b}_j-\vec{b}_k$ and finite pulsewidths.  
To see the impact of these terms, we find that
$$
\vec{S}_1\times\vec{S}_2+\vec{S}_2\times\vec{S}_3+\vec{S}_3\times\vec{S}_1 = 4(\vec{S}_1\times\vec{S}_2\cdot\vec{S}_3) \vec{S} = \sqrt{3}\sigma^y \vec{S},
$$
where $\sigma^y$ is the Pauli-$y$ operator on the $S_{12}$ degree of freedom of the encoded qubit~\cite{Burkard2021}.  Again $\vec{S}$ impacts only $m$, and so this remaining term is proportional strictly to $\sigma^y$ in the encoded subspace and, remarkably, does not cause leakage.   Hence, ignoring gauge and overall phase, effectively
\begin{equation}
	{\bar{H}}^{(0)}\propto \sigma^y.
\end{equation}

\com{
The third line is a consequence of a finite field gradient, $\vec{G}$, causing evolution during pulses of finite width, $\tp$.
To evaluate this term, we must now consider the possible vector directions of our gradient fields $\vec{b}_j$.
If these local gradients point in different spatial directions, then this last first-order noise term could enable leakage.  However, in practice there will typically be a global magnetic field present, either due to an applied field, or due to the magnetism of the earth. 
We take this global field to be along the $z$ direction.
Local gradients not along $\hat{z}$ will, in the presence of fast Larmor precession of electron spins along this direction, average out on the Larmor timescale, and may be neglected.
In a regime where the period of pulsing is comparable to the Larmor period, however, such terms are of relevance, and are addressed more thoroughly in Appendix~\ref{app:filter_function}.
For now, let us consider the case of an applied field sufficiently large to render Larmor precession much faster than pulsing speed, and consider only the $z$-component of the gradient $\vec{G}$.
In this simple case, the result of collinear gradients during finite pulse durations is a coupling to the operator
\begin{equation}
\hat{z}\cdot[(\vec{S}_1+\vec{S}_2)\times\vec{S}_3-\vec{S}_1\times\vec{S}_3]
=\sqrt{3}im\biggl(\ketbra{1}{0}-\ketbra{1}{0}\biggr),
\end{equation}
that is, to $\sqrt{3}m$ times $\sigma^y$ for the DFS qubit with total spin projection $m$.
It is a remarkable feature of this sequence that although collinear magnetic field gradients generally cause leakage without decoupling, NZ1 sequences refocus this leakage to first order.
Since the first two lines of $\bar{H}^{(0)}$ above are effectively identity in the DFS, and the last is proportional to $\sigma^y$, we may summarize our average Hamiltonian omitting overall phase to
\begin{equation}
	\bar{H}^{(0)}=\frac{\sqrt{3}m}{2\pi}G^z\sigma^y.
\end{equation}
}

When initializing $\ket{0}$, the $\sigma^y$ dependence results in oscillations proportional to $\tp$ and to field gradients, which, due to nuclear dynamics, undergo slow fluctuations, resulting in an apparent decay.  
These oscillations could be refocused using higher-order sequences such as the NZ2 sequence $NZNZNZZNZNZN$~\cite{West2012}. Here we employ an alternative method, similar to the reduction of static pulsing errors in CPMG~\cite{Meiboom1958}: we choose to prepare our qubit along the $\hat{y}$-axis of the Bloch sphere such that $\bar{H}^{(0)}$ causes no evolution except on the ignored $m$ degree of freedom.
We do this by including composite single-DFS-qubit rotations so that the evolution becomes, ignoring gauge and overall phase, effectively
\begin{equation}
	U(nT)\approx \mathbb{H} \mathbb{S}^\dag e^{-i\bar{H}^{(0)} nT} \mathbb{S} \mathbb{H} \propto e^{-i\sigma^z n T},
	\label{eq:nzy}
\end{equation}
where $\mathbb{S} = \text{diag}(1,i)$ in the qubit basis and $\mathbb{H}$ is the Hadamard gate (see Ref.~\onlinecite{Andrews2019} for calibration and construction of these single-qubit gates).  
This operator is effectively identity on the prepared and measured singlet $\ket{0,S_{12};m}$ state, regardless of any first-order static field gradient or exchange angle offset.
Since this $NZ$ sequence is first order and applied in the $y$-basis, we refer to this construction as NZ1y.

\storyboard{(Calculate the filter function for NZ1y)}

\begin{figure}[th!]
	\includegraphics[width=\columnwidth]{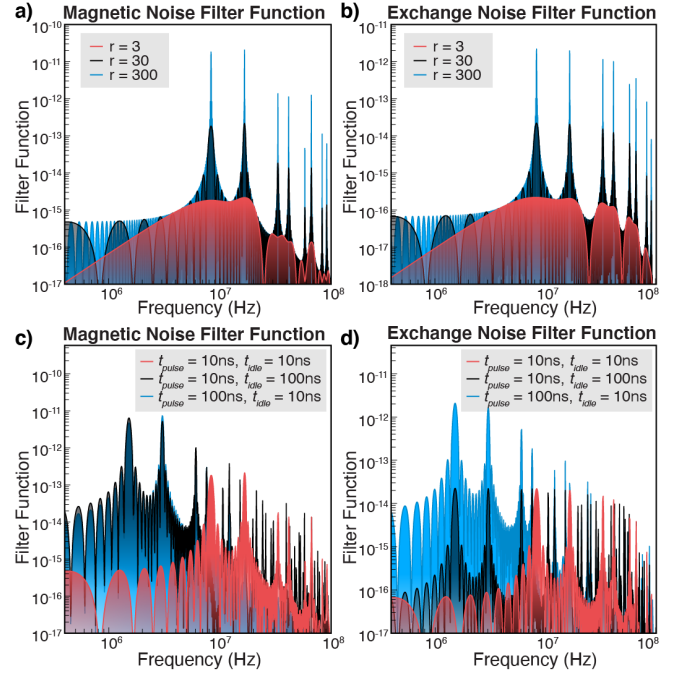}
	\caption{
		Filter functions of the NZ1y sequence construction. 
		\textbf{(a)} and \textbf{(b)} show the magnetic and exchange noise filter functions for a $(NZ)^{r}$ sequence with $\ts{t}{pulse}$= 10 ns and $\ts{t}{idle}$ = 10 ns, and different sequence lengths $r=3,30,300$. 
		\textbf{(c)} and \textbf{(d)} show the magnetic and exchange nosie filter functions for $r=30$ at several different durations $\tp$ and $\ti$. 
	}
	\label{fig:FF}
\end{figure}

The analysis thus far has considered the lowest order average Hamiltonian only, and has found that both static magnetic and exchange noise terms vanish entirely to first order.
We may find the impact of static noise at higher orders by analyzing higher orders of the Magnus expansion.
In doing so, we find that static exchange error angles $\delta\theta_n$ and $\delta\theta_z$  contribute in the next order when initializing along $\hat{z}$ (NZ1z), showing a coherent error of size $(3/32)(\delta \theta_{n}^2+\delta \theta_{z}^2-4\delta\theta_z\theta_n)^2$.  
When initializing along $\hat{y}$ (NZ1y), these errors are canceled even at second order, and to third order:
\begin{multline}
	\text{Miscalibration Error}\approx\\
	\frac{1}{32}
	(\delta \theta_{n}^2+\delta \theta_{z}^2-4\delta\theta_z\theta_n)^2
	(\delta \theta_{n}^2+\delta \theta_{z}^2- \delta\theta_z\theta_n).
\end{multline}
The resilience of NZ1y to static miscalibration error makes it especially effective at exposing noise at higher frequencies.

The analysis of evolution of NZ1y in the presence of high-frequency magnetic and exchange noise is not well handled by average Hamiltonian theory; instead we employ a filter function formalism derived in Appendix~\ref{app:filter_function}.  This finds, for ensemble average $\langle\cdot\rangle$,
\begin{multline}
	1-\langle |\!\!\bra{0}\!U(nT)\!\ket{0}\!\!|^2\rangle  \approx 
 \\
		      \int_0^\infty  d\nu\ S_B(\nu) \mathcal{F}_\mathcal{M}(\nu,nT) 
	        + \int_0^\infty d\nu\ S_E(\nu) \mathcal{F}_E(\nu,nT),
\end{multline}
where $S_B(\nu)$ is the single-sided noise power spectral density (PSD) for magnetic noise $\vec{b}_j(t)$ assumed independent and identical on each dot, and $S_E(\nu)$ is the PSD for exchange noise $\delta J_{jk}(t)/\hbar$, assumed independent and identical for each pair of dots within an exchange axis.
As usual for filter function formalism, the magnetic and exchange filter functions, $\mathcal{F}_\mathcal{M}(\nu,t_n)$ and $\mathcal{F}_E(\nu,t_n)$, respectively, are oscillatory, with a central peak dictated by the frequency of pulsing and with a passband that narrows with the number of pulses.
Note that $\vec{b}_j(t)$ can be composed of non-collinear time-dependent noise, and that $\delta J_{jk}$ can vary from pulse-to-pulse.
In this way, our derivation encapsulates quasi-static noise and white Johnson noise as well as $1/f$ noise. 

Example filter functions are shown in \reffig{fig:FF}.
We plot the magnetic (\reffig{fig:FF}a) and exchange (\reffig{fig:FF}b) noise filter functions for varying number of pulses. 
In both cases, increasing the sequence length does not change the spectral locations of the filter function peaks, but instead changes the sharpness and amplitude of the filter function passbands.
When applying decoupling pulses to a qubit, this analysis predicts an exponential decay in the probability of recovering the initial state upon repeated applications of the decoupling block.
In contrast, changing the total duration of the NZ1 decoupling block (6$\ti$+6$\tp$) shifts the spectral location of the magnetic and exchange filter function peaks, as seen in \reffig{fig:FF}c-d.
The rate of decay of the return probability with the number of decoupling pulses defines the NZ1y error, whose experimental measurement we discuss in the next section.

\section{Experimental Results}\label{sec:results}

\begin{figure}[th!]
	\includegraphics[width=\columnwidth]{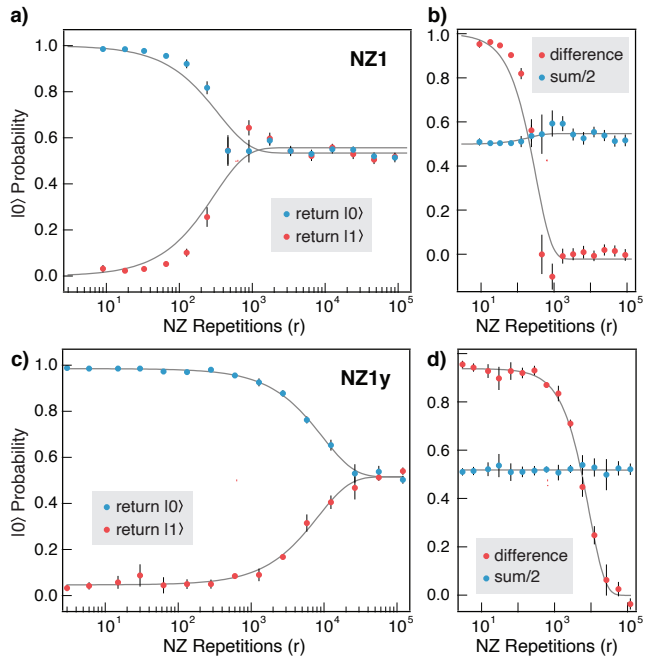}
	\caption{Full-permutation DD of an encoded qubit. 
	\textbf{(a)} NZ1 experiment results, plotting $\ket{0}$  return probability after the decoupling sequence of varying length is applied. 
	The ideal result is either 1 ($\ket{0}$ state, blue points) in half of the runs, or 0 ($\ket{1}$ state, red points) in the other half due to an optionally applied $X$ rotation. 
	\textbf{(b)} Difference and sum curves. 
	\textbf{(c)} NZ1y experiment results, in which the same decoupling sequence is applied to a state that is rotated by $\mathbb{S} \mathbb{H}$ after preparation and $\mathbb{H} \mathbb{S}^\dag$ before measurement, again with both $\ket{0}$ and $\ket{1}$ expected final states. 
	\textbf{(d)} Difference and sum of the NZ1y data traces indicate error within and leakage out of the encoded subspace respectively. 
	A single-exponential fit to the difference curve yields a qubit error rate of $1.8\times10^{-3}$ per pulse for the NZ1 sequence and $5\times10^{-5}$ for the NZ1y sequence, with an indiscernible leakage rate for both sequences. 
	Grey lines are fits to a single exponential constrained to probability $<1$.
	}
	\label{FIG:nzdata}
\end{figure}

\storyboard{(What we are doing to prepare and measure a qubit with these sequences applied)}
We apply the two forms of first-order NZ sequences discussed above to our qubit, with results shown in Fig.~\ref{FIG:nzdata}. 
The experiments begin with initialization of spins to the $\ket{0}$ state and culminate with a measurement of $\ket{0}$ return probability upon ensemble average.
In half of the experimental runs, a final $X$ rotation is applied so that the ideal expected state is $\ket{1}$. 
The sequence employs calibrated voltage pulses of duration $\tp = 10~$ns with time between pulses $\ti = 10~$ns. 
Details of initialization, measurement, and calibration methods are identical to those described in Ref.~\onlinecite{Andrews2019}.
For the qubit in this work, a prepared state decays in $T_2^*=2~\mu$s when device voltages are held in idle and no decoupling sequence is applied. 
 
\storyboard{(NZ experiment method and analysis)}
The first experiment aims to measure the NZ1 error by observing the decay in fidelity when the qubit is subject to repeated applications of the decoupling block.
Fig.~\ref{FIG:nzdata}a shows the $\ket{0}$ and $\ket{1}$ return probabilities after $(NZ)^r$ pulses for $r=0\pmod{3}$.
The condition that $r=0\pmod{3}$ forces the total pulse sequence to be an integer multiple of the base NZ1 decoupling block and, by definition, multiplies the error of a single NZ1 block by $r$.
In order to extract both the error rate in the encoded space and the leakage to the $\ket{Q}$ state, we apply the same methodology as in ``blind" randomized benchmarking~\cite{Andrews2019}, analyzing the sum and difference of the data traces for experiments with expected final states of $\ket{0}$ and $\ket{1}$.
This is shown in Fig.~\ref{FIG:nzdata}b, where we find that the NZ1 sequence results in decay to 1/e in $r=281$, which equates to an infidelity of $1.8\times10^{-3}$ per exchange pulse, and decay time of $T_2=11~\mu$s. 
We attribute the extracted error per pulse to magnetic noise of collinear magnetic gradients that is accumulated during the finite-duration pulses. 
Since these magnetic errors are effectively rotation errors about the $\hat{y}$-axis, as shown in Sec.~\ref{sec:DD}, they can be suppressed by initializing and measuring along $\hat{y}$, as we will now show.

\storyboard{(NZ1y experiment method and results)}  
Next, we apply the NZ1y sequence to the same qubit using the construction  $  \mathbb{S} \mathbb{H} (NZ)^r \mathbb{H} \mathbb{S}^\dag$ and sweeping $r$ with $r=0\pmod{3}$  (Fig.~\ref{FIG:nzdata}c-d).
We find that the NZ1y sequence maintains qubit coherence for 18,000 exchange pulses, equating to an infidelity of $5\times10^{-5}$ per exchange pulse.
The decay time $T_2 = 360~\mu$s is about 32 times longer than the that of the NZ1 sequence with the same pulse and idle durations. 
Additionally, as demonstrated by the sum curve, the NZ1y sequence does not exhibit any leakage into the $\ket{Q}$ state even at over 20,000 exchange pulses, far after the encoded qubit has decohered.
In comparison, blind randomized benchmarking (RB) performed on this qubit exhibited an error rate 64 times higher ($3.5\times10^{-3}$ per Clifford gate), with magnetic noise contributing to half of the total error and a leakage error of $1.7\times10^{-3}$~\cite{Andrews2019}.

\begin{figure}[th!]
	\includegraphics[width=\columnwidth]{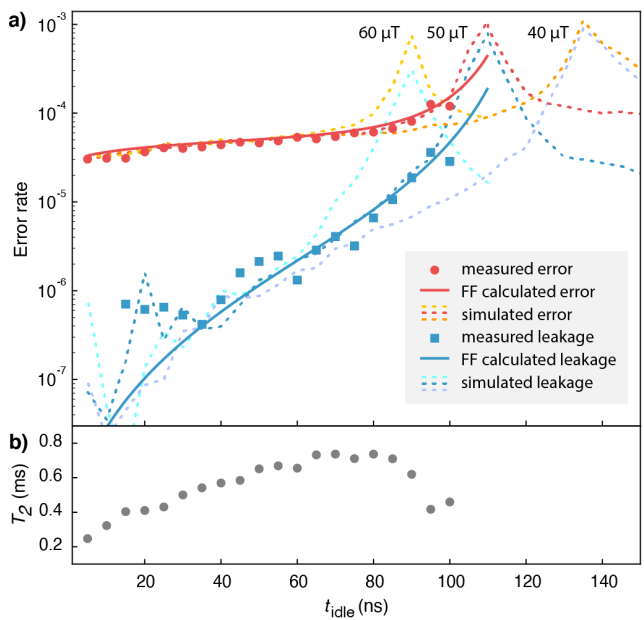}
	\caption{NZ1y sequence with variable idle time. \textbf{(a)} Measured, simulated, and calculated error and leakage per pulse as a function of swept $\ts{t}{idle}$. Simulation results (dotted lines) are shown at three different values of $B_0^z$  to show the effect of Larmor resonance. An analytical calculation (solid line) integrating the filter function with the assumed noise spectra is perfomed at $B_0^z=45~\mu$T to best match the experimental data.  \textbf{(b)} Extracted $T_2$ decay time for each measured sequence reveals a maximum  $T_2 = 720~\mu$s at $\ts{t}{idle}=80~$ns.
	}
	\label{fig:nz_idle}
\end{figure}

\storyboard{(We can increase magnetic error contribution by extending idle time.)}
These results highlight how noise sources couple to RB and the NZ1y sequence at very different rates, allowing the latter to be a distinct, sensitive and targeted probe of error contributions. 
In particular, we will now study the effects of increasing the magnetic error contribution by increasing the idle duration in the NZ1y sequence.
Although the pass-band locations of both the magnetic and exchange noise filter functions depend on the total NZ1 block duration, the magnitude of the exchange noise filter function is insensitive to $\ti$.
This stems from the large on/off ratios achieved with the exchange interaction, with the result that the qubit suffers exchange noise from charge fluctuations only during $\tp$.
Examining Fig.~\ref{fig:FF}c-d, we see that both the magnetic and exchange noise filter functions have peaks in frequency space that are determined by the total $\tp+\ti$, and that an increase in either $\tp$ or $\ti$ is accompanied by an increase in magnitude of the magnetic noise filter function, whereas the exchange noise filter function amplitude is very sensitive to $\tp$ only.
Thus, sweeping $\ti$ can dramatically change the magnetic noise contribution while having a smaller effect on the exchange noise contribution. 

\storyboard{(doing that experiment and looking at the results)}
To this end, we perform a suite of NZ1y experiments, varying  $\ti$ from $5$~ns to $100$~ns while keeping the pulse duration constant at $\tp=10$~ns.
In doing so, we effectively sweep the primary NZ1y passband from 11~MHz to 1.5~MHz, referring to the filter functions shown in Fig.~\ref{fig:FF}c-d.
For each $\ti$ we perform the experiment as in Fig.~\ref{FIG:nzdata}d, extracting both the encoded error and leakage error rates.
The resulting rates are plotted in Fig.~\ref{fig:nz_idle}a, and show that the error rate increases by about a factor of five over the range of the sweep.
At short $\ti$ the leakage error is less than 1/30 of the total error. 
At $\ti=100$~ns, leakage error is significantly higher and magnetic noise contributes to 30\% of the total error rate.
In Fig.~\ref{fig:nz_idle}b we show the total decay time across the range, calculated as $T_2 = (\tp+\ti)/\epsilon$. 
The arc in this plot elucidates a cross-over between an exchange-noise dominated regime at short idle times and a magnetic-noise dominated regime at long idle times, where the filter function extends to include lower frequencies. 
We find a maximum $T_2 = 720~\mu$s at $\ti=80~$ns.

\storyboard{(We simulate this experiment too, and see good agreement.)}
To complement our experimental results, we also performed time-domain Monte-Carlo simulations of these NZ1y experiments across the range of varying $\ti$. 
The simulation includes a constant global magnetic field, $B_0^z$, and two noise sources: 
Magnetic noise~\cite{Eng2015,Kerckhoff2021,Madzik2020} is composed of randomly oriented classical effective vector fields (nuclear polarization $\vec{b}_j$ of  Sec.~\ref{sec:DD}, \refeq{eq:HB}) drawn from a $1/f$ power spectral density out to 10~kHz, beyond which it rolls-off to $1/f^2$. 
Charge noise is simulated by relative exchange fluctuations $\delta J_{jk}(t)/J_{jk}$ drawn from a 1/$f$ distribution with high-frequency roll-off at $> 1$~GHz.  
For both noise sources, low-frequency cut-offs to noise are selected to be lower than the inverse of the full experiment averaging time.  
The noise source amplitudes are calibrated via simulation of free-evolution~\cite{Eng2015} and triple-dot Rabi~\cite{Reed2016} experiments to reproduce the measured qubit parameters of $T_2^*=2~\mu$s and 25 Rabi oscillations at $1/e$.
Finally, we assume a global magnetic field $B_0^z$  near the approximate value of Earth's field in Malibu, CA.
The experimental data shows that both error and leakage increase near the Larmor resonance condition, and our simulation and filter function calculation captures this effect (the average Hamiltonian analysis of Sec.~\ref{sec:DD} does not). 
The simulation results for several field values are shown as dashed lines in Fig.~\ref{fig:nz_idle} and have good correspondence with the experimental data for the case of $B_0^z=50~\mu$T.
An increase in measured error seen near $\ti=95~$ns is ascribed to pulsing near resonance with the electron Larmor frequency, $g\mu_B B_0^z/h$.
In addition to the simulation, an analytical calculation (solid line) integrating the filter function with the assumed noise spectra was also performed, with best match to the experimental data found at $B_0^z=45~\mu$T.  The slight discrepancy in best-fit field is likely due to the approximate character of the filter-function analysis, especially near a resonance-like condition.
We note that the extended coherence of this sequence allows validation of our noisy exchange simulator (and its assumed noise spectra) to greater fidelity than was accessible by RB~\cite{Andrews2019}.

\storyboard{(Conclusion)}
We have experimentally demonstrated the full-permutation dynamical decoupling sequence in a triple-dot, exchange-only qubit. 
The NZ1y sequence features an exceptionally low error rate of $5\times10^{-5}$ per pulse when applied to our qubit, which is attributed to effective echoing of magnetic noise and suppression of low-frequency fluctuations in both exchange and local magnetic fields.
Additionally, due to strong insensitivity to pulse miscalibration, the sequence allows for the $1/f$ noise tail and any high frequency noise processes to be examined independent of calibration accuracy. 
Our simulations accurately account for the measured error and leakage rates in this experiment.
We have studied how extending $\ti$ increases both error within and leakage out of the DFS by increasing susceptibility to magnetic fluctuations, as is clear from the calculable filter function, and we find agreement with time-domain simulation. 
The NZ1y sequence has proven exceptionally helpful in validating our error model for this qubit, and expect its continued utility as exchange-only qubits improve thanks to advancements such as the SLEDGE architecture~\cite{Ha2022,Weinstein2022}.
Permutation decoupling may be similarly valuable for validating error models in other platforms as well, especially with increased control fidelities for swapping qubits.

\acknowledgements{
We thank John Carpenter for assistance with all figures.
We acknowledge Cody Jones and Tyler Keating for significant technical contributions.
}

\appendix

\section{Sequence Filter Functions}
\label{app:filter_function}
%This is the appendix for the NZ filter functions
\newcommand{\bs}{\bm\sigma}
In this appendix, we compute filter functions for the NZ sequences discussed in the main text. Filter functions are computed in first-order Magnus-expansion perturbation theory, which is sufficient to compute state preservation and orthogonal state transition probabilities to second order in the noise. 

% Here is where Thaddeus steps in and makes the appendix looks like part of the rest of the document . . . on 3/30/2022.  (Took COVID to get him the time . . .)

\com{  % I am commenting this out as we already delivered this Hamiltonian, 
	   % and there is not actually a good reason to introduce a new notation here.
The zeroth order triple dot Hamiltonian is given by
\begin{eqnarray}
H_0(t) &=&\frac{1}{2} B_0 \left( \sigma_3^{(1)} + \sigma_3^{(2)} + \sigma_3^{(3)}\right) \nonumber\\
&+& \frac{1}{4} J_0^{(1)}(t) \bs^{(2)}.\bs^{(3)} 
+ \frac{1}{4} J_0^{(3)}(t) \bs^{(1)}.\bs^{(2)} .
\label{eq:H0}
\end{eqnarray}
In this appendix, Latin subscripts refer to spatial directions, and Greek superscripts refer to dot or spin labels, i.e., $\sigma_i^{(\alpha)}$ is the Pauli matrix in direction $i$ for spin $\alpha$. Zero subscripts refer to noiseless quantities. The first group of terms in the Hamiltonian correspond to a global constant uniform magnetic field in the $z$ direction, while the remaining terms are the time-dependent noiseless exchange interactions on the encoded qubit $N$ and $Z$ axes. Superscripts for the exchange strength $J_0^{(\alpha)}$ are given in ``complementary" or dual notation, so that $J_0^{(1)}$ refers to the exchange strength between dots 2 and 3 (qubit $N$ axis) and $J_0^{(3)}$ refers to the exchange strength between dots 1 and 2 (qubit $Z$ axis); $J_0^{(2)}$ between dots 1 and 3 is always zero. The zeroth order noiseless exchange interactions $J_0^{(1)}, J_0^{(3)}$ are assumed to be perfect square pulses---either ``on'' or ``off"---and that $N$ and $Z$ exchanges are never on simultaneously. To ease notational bloat, in the following we drop parentheses in superscripts.
The perturbative Hamiltonian containing noise terms is given by
\begin{equation}
H_1(t) = \frac{1}{2} \sigma_i^\alpha B_i^\alpha(t) + \frac{1}{4} E^\alpha J^\alpha(t),
\label{eq:H1}
\end{equation}
with summation over repeated indices implied, and the exchange operator defined as
\begin{equation}
E^{\{1,2,3\}} \equiv \{ \bs^2.\bs^3, \bs^1.\bs^3, \bs^1.\bs^2 \}.
\label{eq:Eops}
\end{equation}
$B_i^\alpha(t)$ and $J^\alpha(t)$ are magnetic and exchange noise, respectively, with the exchange noise only present when the corresponding zeroth order exchange gate is active. 

With the above definitions of $H_0$ and $H_1$, first order interaction picture perturbation theory gives a time-integrated interaction Hamiltonian of
\begin{equation}
\int_0^t ds\ H_I(s) = \frac{1}{2} \sigma_i^\alpha \theta_i^\alpha(t) - \frac{1}{4} \epsilon_{ilm} D^\alpha_{lm} \phi_i^\alpha(t) + \frac{1}{4} E^\alpha \psi^\alpha(t),
\label{eq:HI}
\end{equation} 

\begin{equation}
	D_{jk}^{\{1,2,3\}} \equiv \{ \sigma_j^2 \sigma_k^3, \sigma_j^3 \sigma_k^1, \sigma_j^1 \sigma_k^2 \},
	\label{eq:Dops}
\end{equation}
}

We mathematically separate noise by first defining the noiseless Hamiltonian $H_0(t)$ to include $\ts{H}{control}(t)$, as in \refeq{eq:Hcontrol}, under the assumption of perfectly calibrated, non-overlapping square pulses for $J_{12}(t)$ and $J_{23}(t)$, as well a uniform, constant, applied magnetic field, $\vec{B}_0$, explicitly
\begin{equation}
	\label{eq:H0}
	H_0(t) = \ts{H}{control}(t) - g\mu_B \vec{B}_0\cdot\vec{S},
\end{equation}
with total spin $\vec{S}=\sum_j\vec{S}_j$ as before.   We further introduce the shorthand angular Larmor frequency $\nu_0 = |g\mu_B\vec{B}_0|/h$.  As all terms of \refeq{eq:H0} commute, it is straightforward to integrate 
\begin{equation}
	U_0(t) = \exp\left(-i\int_0^t d\tau H_0(\tau)\right),
\end{equation}
which we use to classify terms in the time-integral of the interaction picture noise Hamiltonian $\ts{H}{noise}(t)$ from \refeq{eq:Hnoise} as
% \theta -> u
% \phi -> v
% \psi -> w
\begin{multline}
	\label{eq:HI}
	\Xi(t)=
	\int_0^t d\tau\ U^\dag_0(\tau)\ts{H}{noise}(\tau)U^{\phantom\dag}_0(\tau)
  = \sum_j \vec{S}_j\cdot\vec{u}_j(t) 
\\
  -\sum_{\boldsymbol\alpha} \vec{S}_{\alpha_1}\times\vec{S}_{\alpha_2}\cdot\vec{v}_{\alpha_3}(t) + 
  \vec{S}_{\alpha_1}\cdot\vec{S}_{\alpha_2} w_{\alpha_3}(t).
\end{multline} 
Here we employ the vector of subscripts $\boldsymbol{\alpha}$ to enumerate over the ordered sets of dot indices $\{[1,2,3],[2,3,1],[3,1,2]\}$.   We have thus sorted terms into error phase angle integrals given by
\begin{eqnarray}
\vec{u}_j(t) &\equiv& \int_0^t d\tau\ \sum_k f_{jk}(\tau) \vec{R}(\tau)\cdot \vec{b}_k(\tau),
\label{eq:magnetic_error_phase}\\
\vec{v}_{j}(t) &\equiv& \int_0^t d\tau\ \sum_{k\ell} g_{j k}(\tau) \epsilon_{\ell k m} \vec{R}(\tau)\cdot \vec{b}_m(\tau),
\label{eq:magnetic_error_phase_finite_pulse_width}\\
w_{j}(t) &\equiv & \int_0^t d\tau\ \sum_{\boldsymbol\alpha} \delta J_{\alpha_1\alpha_2}(\tau)h_{j\alpha_3}(\tau) .
\label{eq:exchange_error_phase}
\end{eqnarray}
The error phase angle expressions have straightforward physical explanations: the magnetic error phases $\vec{u}_j(t)$, for example, include external-magnetic-field-generated Larmor precession through the spatial rotation matrix
\begin{equation}
\vec{R}(t) \equiv 
\begin{pmatrix}
\phantom{-}\cos (2\pi\nu_0 t) & \sin (2\pi\nu_0 t) & 0 \\
         - \sin (2\pi\nu_0 t) & \cos (2\pi\nu_0 t) & 0 \\
0 & 0 & 1
\end{pmatrix},
\label{eq:larmor_rotation}
\end{equation}
and $\pi$-pulse induced permutations through the ``switching" matrices $f_{jk}(t)$, which in the zero exchange pulse width limit are natural permutation representation matrices of the symmetric group $S_3$ corresponding to a particular sequence of $N$ and $Z$ pulses. For finite-width exchange pulses, the $f_{\alpha\beta}(t)$ also include transitional matrices giving the time dependence of the mapping of Pauli spin matrices from one dot to another as exchange is applied. The finite exchange pulse width additionally gives rise to a weight-two Pauli term reflected as $\vec{v}_j(t)$, with $g_{ek}(s)$ another set of switching matrices that can be determined from $f_{jk}(s)$ through a masking procedure, and $\epsilon_{jk\ell}$ is the Levi-Cevita tensor. Exchange error phases $w_j(t)$ similarly have switching matrices $h_{jk}(t)$ that are determined from $f_{jk}(t)$.     

\com{
The eight-dimensional Hilbert space of three spins has an angular momentum basis specified by quantum numbers total spin $S$, total $z$ spin projection $m$, and spin of the first two dots $S_{12}$: $\ket{S,m,S_{12}}$, the usual decoherence free subsystem basis \cite{DiVincenzo2000}.
\mtr{you changed the notation from the introduction, where S12 was the first quantum number in the ket.}
$S_{12}=0,1$ give encoded $\ket{0}_E$, $\ket{1}_E$ states in the $S=1/2$ subspace, and $m=\pm1/2$ corresponds to a gauge qubit.}
Utilizing our previously introduced basis notation $\ket{S_{12},S;m}$, we enumerate the eight-dependent basis as follows.  The encoded states (duplicated twice due to gauge freedom,) are the first four states:
% rho -> cos theta/2
% eta -> phi
\begin{eqnarray*}
\ket{1} & \equiv & \cos\frac\theta{2} \ket{0,1/2;+1/2} + \sin\frac\theta{2} e^{i \phi} \ket{1,1/2;+1/2}, \label{eq:state1} \\
\ket{2} & \equiv & \cos\frac\theta{2} \ket{0,1/2;-1/2} + \sin\frac\theta{2} e^{i \phi} \ket{1,1/2;-1/2}, \label{eq:state2}  \\
\ket{3} & \equiv & \sin\frac\theta{2} \ket{0,1/2;+1/2} - \cos\frac\theta{2} e^{i \phi} \ket{1,1/2;+1/2}, \label{eq:state3}  \\
\ket{4} & \equiv & \sin\frac\theta{2} \ket{0,1/2;-1/2} - \cos\frac\theta{2} e^{i \phi} \ket{1,1/2;-1/2}, \label{eq:state4}
\end{eqnarray*}
where $\theta$ and $\phi$ specify the Bloch-sphere-polar angles of the encoded state, which we will alter according to the experiment we choose to analyze. 
The remaining four states are the leakage quadruplet,
\begin{eqnarray*}
\ket{5} & \equiv & \ket{1,3/2;+3/2}, \label{eq:state5} \\
\ket{6} & \equiv & \ket{1,3/2;+1/2}, \label{eq:state6} \\
\ket{7} & \equiv & \ket{1,3/2;-1/2}, \label{eq:state7} \\
\ket{8} & \equiv & \ket{1,3/2;-3/2}, \label{eq:state8}
\end{eqnarray*}
Assuming that the initial encoded state is totally mixed in gauge, all probabilities of interest can be written as
\begin{equation}
P(t) = \frac{1}{2} \biggl\langle \sum_f \sum_{i=1}^2  \left|\bra{f} U_0(t)
	\ts{\tilde{U}}{noise}(t)\ket{i} \right|^2  \biggr\rangle,
\label{eq:general_probabilities}
\end{equation}
where $\ts{\tilde{U}}{noise}(t)$ is the interaction propagator for $\ts{H}{noise}(t)$ under $H_0(t)$, and the outer brackets $\langle \cdot\rangle$ refer to ensemble averaging over noise terms.  We define the ``state preservation probability" $P_{I}(t)$ for which the final states to be summed over are $\ket{f}\in\{\ket{1},\ket{2}\}$; we also define the ``encoded error probability" for which $\ket{f}\in\{\ket{3},\ket{4}\}$. The ``leakage error probability" would have $\ket{f}\in\{\ket{5},\ket{6},\ket{7},\ket{8}\}$.  Assuming that the sequence of $N$ and $Z$ pulses results in the identity, $U_0(t)$ acts trivially to the left on $\bra{f}$ states, so that all probabilities depend only on the interaction propagator. Therefore, to lowest order in the Magnus expansion, all probabilities take the form
\begin{equation}
P(t) \approx \biggl\langle \sum_{i,f} \bigl|
	\bigl\langle f \bigl| \bigl[1-i\Xi(t) - \frac{1}{2} \Xi^2(t) \bigr]
				    \bigr| i \bigr\rangle \bigr|^2 \biggr\rangle
\label{eq:approximate_probabilities}
\end{equation}
which is correct to second order in the noise.
\com{
\begin{equation}
P(t) \approx \biggl\langle 
	\frac{1}{2} \sum_{i,f} \biggl| 
		\biggl\langle f \bigl|
			\exp\bigl[-i \int_0^t ds\ H_I(s)\bigr]\bigr|i\bigr\rangle \biggr|^2 \biggr\rangle_\mathrm{noise} 
\approx \left \langle \sum_{i,f} \left| \bbraket{f\bigg| 1- i \int_0^t ds\ H_I(s) - \frac{1}{2} \left( \int_0^t ds\ H_I(s) \right)^2}{i} \right|^2 \right\rangle_\mathrm{noise},
\label{eq:approximate_probabilities}
\end{equation}
}
Substituting Eq.~(\ref{eq:HI}) into Eq.~(\ref{eq:approximate_probabilities}), with the assumptions that the noise is zero mean and that magnetic and exchange errors are uncorrelated, we arrive at quadratic terms of the form
\com{
\begin{equation}
\mathcal{Q} = 
Q_{\mathcal{M},11}^{\alpha\beta} \left\langle \theta_i^\alpha \theta_i^\beta \right\rangle +
Q_{\mathcal{M},12}^{\alpha\beta} \left\langle \theta_i^\alpha \phi_i^\beta \right\rangle 
+Q_{\mathcal{M},21}^{\alpha\beta} \left\langle \phi_i^\alpha \theta_i^\beta \right\rangle +
Q_{\mathcal{M},22}^{\alpha\beta} \left\langle \phi_i^\alpha \phi_i^\beta \right\rangle +
Q_{\mathcal{E}}^{\alpha\beta} \left\langle \psi^\alpha \psi^\beta \right\rangle,
\label{eq:quadratic_angle_form}
\end{equation}
}
\begin{align}
	\mathcal{Q} = 
	\sum_{jk}
	&Q_{\mathcal{M},jk}^{(11)}(\theta,\phi)
		\left\langle \vec{u}_j\cdot\vec{u}_k \right\rangle +
	Q_{\mathcal{M},jk}^{(12)}(\theta,\phi)
		\left\langle \vec{u}_j\cdot\vec{v}_k\right\rangle 
\notag\\
	&Q_{\mathcal{M},jk}^{(21)}(\theta,\phi)
		\left\langle \vec{v}_j\cdot\vec{u}_k\right\rangle +
	Q_{\mathcal{M},jk}^{(22)}(\theta,\phi)
		\left\langle \vec{v}_j\cdot\vec{v}_k\right\rangle 
\notag\\
	&\hspace{1in}
	+Q_{\mathcal{E},jk}(\theta,\phi)
		\left\langle w_j w_k \right\rangle,
	\label{eq:quadratic_angle_form}
\end{align}
where the error phase angles are given in Eqs.~(\ref{eq:magnetic_error_phase}--\ref{eq:exchange_error_phase}), and the angle brackets are noise averages. The $Q$ matrices are dependent on the encoded initial state parameters $\theta$ and $\phi$, and differ for each of the three probabilities of interest. For state preservation probability, $P_S = 1 - \mathcal{Q}$; for both encoded and leakage error probabilities, $P_{E,L} = \mathcal{Q}$, with suitable substitutions for the different $Q$ matrices. The quadratic magnetic-error-phase-angle noise average, under the assumption that random noise fields at different dots and along different axes are uncorrelated and identically distributed with correlation function $\langle b(\tau_1) b(\tau_2)\rangle$, is
\com{
\begin{eqnarray}
\left\langle \theta_i^\alpha \theta_i^\beta \right\rangle &=& \int_0^t ds_1 \int_0^t ds_2\ R_{ij}(s_1) R_{ik}(s_2) f^{\alpha\gamma}(s_1) f^{\beta\mu}(s_2)
\left\langle B_j^\gamma(s_1) B_k^\mu(s_2) \right\rangle \\
&=&  \int_0^t ds_1 \int_0^t ds_2\ R_{ij}(s_1) R_{ik}(s_2) f^{\alpha\gamma}(s_1) f^{\beta\mu}(s_2) \delta^{\gamma\mu} \delta_{jk} C_B(s_1-s_2) \\
&=&  \int_0^t ds_1 \int_0^t ds_2\ R_{ij}(s_1) R_{ik}(s_2) f^{\alpha\gamma}(s_1) f^{\beta\gamma}(s_2) \int_0^\infty d\nu S_B(\nu) \cos 2\pi\nu (s_1 - s_2).
\end{eqnarray}
}
\begin{widetext}
\begin{align}
	\langle \vec{u}_j \cdot \vec{u}_k \rangle &= 
     \int_0^t d\tau_1 \int_0^t d\tau_2\ \sum_{\ell m}
     	f_{j\ell}(\tau_1) f_{km}(\tau_2)
	\left\langle \vec{b}_\ell(\tau_1) \cdot\vec{R}^T(\tau_1)
	                                  \cdot \vec{R}(\tau_2)\cdot\vec{b}_m(\tau_2) \right\rangle \\
	&=  
	\int_0^t d\tau_1 \int_0^t d\tau_2\ \text{Tr}[\vec{R}^T(\tau_1)\vec{R}(\tau_2)] \langle b(\tau_1) b(\tau_2)\rangle \sum_\ell f_{j\ell}(\tau_1) f_{k\ell}(\tau_2) \\
	&=  \int_0^t d\tau_1 \int_0^t d\tau_2\ 
		[1+2\cos(2\pi\nu_0(\tau_1-\tau_2))]
		\int_0^\infty d\nu S_B(\nu) \cos[2\pi\nu (\tau_1 - \tau_2)] \sum_\ell f_{j\ell}(\tau_1) f_{k\ell}(\tau_2).
\end{align}
The final equality uses the Wiener-Khinchin theorem, with $S_B(\nu)$ the one-sided magnetic noise power spectral density for any component $\vec{b}_j(t)$ in any dot.  Now define the Fourier-like transform of the switching matrices
\begin{equation}
\tilde{f}_{jk}(\nu,t) \equiv \int_0^t d\tau\ e^{-2\pi i \nu \tau} f_{jk}(\tau).
\label{eq:switching_matrix_ft}
\end{equation}
Performing the sum over the Larmor rotation matrices and using the symmetry property of the $Q_{\mathcal{M},jk}^{(11)}$ matrix allows us to write
\begin{multline}
Q_{\mathcal{M},jk}^{(11)} 
	\left\langle\boldsymbol\theta_j\cdot\boldsymbol\theta_k\right\rangle 
=
\\
Q_{\mathcal{M},jk}^{(11)} \int_0^\infty d\nu\  S_B(\nu)  \sum_\ell
\bigl[ 
	\tilde{f}_{j\ell}\left(\nu,t\right) 
	\tilde{f}_{k\ell}^*\left(\nu,t\right)
   +\tilde{f}_{j\ell}\left(\nu+\nu_0,t\right) 
	\tilde{f}_{k\ell}^*\left(\nu+\nu_0,t\right) +
   +\tilde{f}_{j\ell}\left(\nu-\nu_0,t\right)   
    \tilde{f}_{k\ell}^*\left(\nu-\nu_0,t\right)
\bigr].
\label{eq:quadratic_magnetic_form}
\end{multline}
All terms to the right of $S_B(\nu)$ are the (weight-one Pauli) magnetic error phase contribution to the magnetic noise filter function. As with other filter function analyses \cite{Kerckhoff2021}, the filter function is the squared magnitude of the Fourier transform of a switching function, with geometric factors arising from the initial state. The finite external magnetic field introduces sideband contributions to the filter function; in the limit of large field these contributions go to zero. The remaining quadratic-error phase-average terms in Eq.~(\ref{eq:quadratic_angle_form}) all have the same structure as Eq.~(\ref{eq:quadratic_magnetic_form}) with different quadratic combinations of $\tilde{f}$, $\tilde{g}$, and $\tilde{h}$, and include the effects of weight-two Pauli finite-pulse-width magnetic errors as well as exchange errors.

Combining the algebra and definitions above, we now summarize the total approximate state preservation probability, given by 
\begin{equation}
P_S \approx 1 - \int_0^\infty d\nu\ S_B(\nu) \mathcal{F}^\mathcal{I}_\mathcal{M}(\theta,\phi,\nu,M,\tp,\ti) 
- \int_0^\infty d\nu\ S_E(\nu)  \mathcal{F}^\mathcal{I}_E(\theta,\phi,\nu,M,\tp,\ti), 
\end{equation}
where $\mathcal{F}^\mathcal{I}_\mathcal{M}$ is the magnetic noise filter function for state preservation infidelity, dependent on the initial state parameters $\theta$ and $\phi$, as well as the frequency $\nu$, the number of repetitions $M$ of the six-pulse $NZNZNZ$ sequence, and the pulse width $\tp$ and idle time between pulses $\ti$; $S_E(\nu)$ is the exchange noise power spectral density and  $\mathcal{F}^\mathcal{I}_E$ is the exchange noise filter function for state preservation infidelity.   Writing the pulse width and idle time in terms of the fractional pulse width $f=\tp/\tau$ and pulse repetition time $\tau=\tp+\ti$, we have the NZ1y magnetic noise filter function for state preservation infidelity,
\begin{multline}
\mathcal{F}^\mathcal{I}_\mathcal{M}(\pi/2,\pi/2,\nu,M,f\tau,(1-f)\tau) 
=
\frac{1}{6 \pi ^2 \nu ^2 \left(1-4 f^2 \nu ^2 \tau ^2\right)^2}
\frac{\sin^2 6M\pi\nu \tau}{\sin^2 6\pi \nu \tau}
\times
\\
\Bigg\{
7 + 32 f^4 \nu ^4 \tau ^4+16 \left(4 f^2 \nu ^2 \tau ^2-1\right) \sin ^3(\pi  \nu  \tau ) \cos ^2(\pi  \nu  \tau ) 
(2 \cos (4 \pi  \nu  \tau )+1) \sin (2 \pi  f \nu  \tau )\times
\\
 \left(-\sqrt{3} f \nu  \tau
    \sin (\pi  \nu  \tau )+\cos (\pi  \nu  \tau )+2 \cos (3 \pi  \nu  \tau )\right)
\\ 
    +f \nu  \tau  \bigg[f \nu  \tau  \left(2 f \nu  \tau  \left(-16 f \nu  \tau  \cos (12 \pi  \nu  \tau )-4 \sqrt{3} \sin
   ^3(2 \pi  \nu  \tau ) (2 \cos (4 \pi  \nu  \tau )+1)\right)+\cos (2 \pi  \nu  \tau )-8 \cos (4 \pi  \nu  \tau )
\right.
\\
\left.
   +6 \cos (6 \pi  \nu  \tau )-4 \cos (8 \pi  \nu  \tau )-7 \cos (10 \pi  \nu  \tau )+12
   \cos (12 \pi  \nu  \tau )\right)+2 \sqrt{3} \sin ^3(2 \pi  \nu  \tau ) (-16 \cos (2 \pi  \nu  \tau )+6 \cos (4 \pi  \nu  \tau )+7)\bigg]
\\
   +2 \sin ^2(2 \pi  \nu  \tau ) \cos (2 \pi  f \nu  \tau )
   \bigg[f \nu  \tau  \left(2 f \nu  \tau  \left(-2 \sqrt{3} f \nu  \tau  \sin (6 \pi  \nu  \tau )
   +8 \cos (2 \pi  \nu  \tau )+12 \cos (4 \pi  \nu  \tau )+7 \cos (6 \pi  \nu  \tau )
\right.\right.
\\
\left.\left.
   +4 \cos (8 \pi  \nu  \tau )+14\right)+\sqrt{3} (4 \sin (2 \pi  \nu  \tau )-8 \sin (4 \pi  \nu  \tau )+3 \sin (6 \pi  \nu  \tau ))\right)
   -10 \cos (2 \pi  \nu  \tau )-2 \cos (6 \pi  \nu  \tau )-2 \cos (8 \pi  \nu  \tau )+5\bigg]
\\   
%\left.
   -4 \cos (2 \pi  \nu  \tau )-4 \cos (4 \pi  \nu  \tau )+3 \cos (6 \pi  \nu  \tau )-2 \cos (8 \pi  \nu  \tau )
   +\cos (10 \pi  \nu  \tau )-\cos (12 \pi  \nu  \tau )\Bigg\}
\\
+ \mathrm{sideband\ terms},
\end{multline}
where the sideband terms are determined from the displayed expression by the substitutions $\nu \rightarrow \nu \pm \nu_0$.
The NZ1y exchange-noise filter function for state preservation infidelity is
\begin{equation}
\mathcal{F}^\mathcal{I}_E(\pi/2,\pi/2,\nu,M,\tp,\ti) = 
\frac{2(2+\cos 4\pi \nu\tau) \sin^2 \pi\nu\tp \sin^2 2\pi\nu\tau}{\pi^2\nu^2}
\frac{\sin^2 6M\pi\nu \tau}{\sin^2 6\pi \nu \tau},
\label{eq: f_i_e}
\end{equation}
which has no sideband contributions and $\tau=\ti+\tp$. The main (not sideband) NZ1y magnetic noise filter function at DC ($\nu=0$) is zero, consistent with the average Hamiltonian theory result in the main text.
%which can also be determined by multiplying propagators for piecewise constant-in-time Hamiltonians including DC magnetic noise and finite exchange pulse width: the resulting average Hamiltonian after the six-pulse $NZNZNZ$ sequence corresponds to a rotation in the Y direction. 
In general the arbitrary initial state magnetic noise filter function at $\nu=0$ is given by (ignoring sideband terms)
\begin{equation}
\mathcal{F}^\mathcal{I}_\mathcal{M}(\theta,\phi,0,M,\tp,\ti) = 
\frac{18 M^2 \tp^2 \left( \cos^2 \phi + \cos^2\theta \sin^2 \phi\right)}{\pi^2},
\label{eq:zero_frequency_NZ1_arb_state}
\end{equation}
which is only zero at $\pm \hat{y}$ (i.e. $\theta=\pi/2=\pm\phi$) initializations. Initializations in other directions quadratically (coherently) accumulate finite pulse width DC magnetic noise errors under NZ pulsing.

In the limit of zero pulse width, the magnetic noise filter function for NZ1y is
\begin{equation}
\mathcal{F}^\mathcal{I}_\mathcal{M}(\pi/2,\pi/2,\nu,M,0,\ti) = \frac{64 (2+\cos 4\pi\nu\ti)\cos^2 \pi\nu\ti \sin^4 \pi\nu\ti}{3\pi^2\nu^2} \frac{\sin^2 6M\pi\nu \ti}{\sin^2 6\pi \nu \ti} + \mathrm{sideband\ terms}.
\label{eq:zero_width_magnetic_noise_NZ1Y_filter_function}
\end{equation}
The corresponding zero-pulse-width exchange noise filter function is, of course, zero.

The approximate encoded error probability for NZ1y, i.e. the probability of flipping between encoded qubit states (without leakage) is given by
\begin{equation}
P_E = \int_0^\infty d\nu\ S_B(\nu) \mathcal{F}^\mathcal{E}_\mathcal{M}(\theta,\phi,\nu,M,\tp,\ti) 
+ \int_0^\infty d\nu\ S_E(\nu)  \mathcal{F}^\mathcal{I}_E(\theta,\phi,\nu,M,\tp,\ti),
\end{equation}
with the encoded error magnetic noise filter function given by
\begin{multline}
\mathcal{F}^\mathcal{E}_\mathcal{M}(\pi/2,\pi/2,\nu,M,f \tau,(1-f)\tau) = 
\frac{1}{12 \pi ^2 \nu ^2 \left(1-4 f^2 \nu ^2 \tau ^2\right)^2}
\frac{\sin^2 6M\pi\nu \tau}{\sin^2 6\pi \nu \tau} \times
\\
\Bigg[
7+
32 f^4 \nu ^4 \tau ^4
-32 f^4 \nu ^4 \tau ^4 \cos (12 \pi  \nu  \tau )
-12 f^2 \nu ^2 \tau ^2
+6 f^2 \nu ^2 \tau ^2 (\cos (2 \pi  \nu  \tau )-\cos (10 \pi  \nu  \tau )+2 \cos (12 \pi  \nu  \tau ))
\\
+2 \sin ^2(2 \pi  \nu  \tau ) \bigg\{ \left(4 f^2 \nu ^2 \tau ^2-1\right) (2 \sin (4 \pi  \nu  \tau )-\sin (6 \pi  \nu  \tau )+2 \sin (8 \pi  \nu  \tau )) \sin (2 \pi  f \nu  \tau )
\\
+\big(4 f^2 \nu ^2 \tau^2 (6 \cos (2 \pi  \nu  \tau )+4 \cos (4 \pi  \nu  \tau )+3 \cos (6 \pi  \nu  \tau )+2 \cos (8 \pi  \nu  \tau )+3)
\\   
   -10 \cos (2 \pi  \nu  \tau )-2 \cos (6 \pi  \nu  \tau )-2 \cos (8 \pi  \nu  \tau )+5\big) \cos (2 \pi  f \nu  \tau )\bigg\}
\\
-4 \cos (2 \pi  \nu  \tau )-4 \cos (4 \pi  \nu  \tau )+3 \cos (6 \pi  \nu  \tau )-2 \cos (8 \pi  \nu  \tau )+\cos (10 \pi  \nu  \tau )-\cos (12 \pi  \nu  \tau )
\Bigg]
\\
+ \mathrm{sideband\ terms}.
\end{multline}
Notice that the encoded error exchange noise filter function is the same as the exchange noise filter function for state preservation infidelity since exchange errors do not cause leakage. The remaining leakage error magnetic noise filter function is the difference of the state preservation infidelity and encoded error filter functions.

\section{Non-rectangular pulses}
\label{app:nonsquare_pulses}
% This is the appendix for NZ1y charge noise response with non-square pulses

We now discuss the effect of non-rectangular exchange pulses on the NZ filter function response.
For simplicity we consider here the case of exchange noise, $\mathcal{F}^\mathcal{I}_\mathcal{E}$.
Non-rectangular pulse shapes correspond to varying noise sensitivity over the duration of the exchange pulse, which can qualitatively modify the filter function envelope.
The exchange pulse shape enters into the filter function expression of \refeq{eq: f_i_e} via the switching terms $h_{jk}(t)$ defined in \refeq{eq:exchange_error_phase}, which encode the time-dependent mapping of Pauli spin matrices over the NZ sequence.
Assuming all pulses have the same shape, the integral over these switching terms is separable into two components corresponding to the pulse shape and to the pulses' relative time shifts over the NZ sequence.
Specific to NZ1y, we find that 
\begin{equation}
\mathcal{F}^\mathcal{I}_\mathcal{E} =
2 \left| h_\mathrm{pulse} \left(\nu, \tp \right) \right|^2 \left( 2+ \cos 4\pi \nu \tau \right) \sin^2 2 \pi \nu \tau
\frac{\sin^2 6M\pi\nu \tau}{\sin^2 6\pi \nu \tau}.
\end{equation}
\end{widetext}

For rectangular pulse with duration $\tp$,  
\begin{equation}
h_{\textrm{rect}} \left( \nu, \tp \right)= \frac{i}{2\pi\nu}\left(e^{-i 2 \pi \nu \tp} - 1\right)
\end{equation}
and we recover the results of \refeq{eq: f_i_e}.  
For a trapezoidal voltage pulse with ramp time $t_R$, full-width-half-maximum $\tp$, on-off ratio $\alpha$ between minimum and maximum sensitivity values and an exponential scaling between exchange energy and voltage, 
\begin{multline}
h_{\textrm{trapz}}\left( \nu, \tp \right) =  
	\frac{1}{\alpha} \frac{i}{2 \pi (\nu + i \, \nu_R)}\left( 
		e^{-i 2 \pi \left( \nu + i \, \nu_R \right) t_R} - 1 
		\right) 
\\
+ \frac{i}{2 \pi \nu}\left(e^{-i 2 \pi \nu \tp } -  e^{- i 2 \pi \nu t_R} \right) 
\\
+ \frac{i e^{-i 2 \pi \nu \tp} }{2 \pi (\nu - i \, \nu_R)}
	\left( e^{-i 2 \pi (\nu - i \, \nu_R)t_R} - 1 \right),
\end{multline}
where $2 \pi \nu_R = \left(1 - 1/\alpha\right) / t_R$.
Compared to rectangular pulses, this filter envelope has a higher frequency cut-off but with a faster decay rate.

\bibstyle{revbst}
\bibliography{nz_paper}

\end{document}